\documentclass[aps,prb,array,epsfig,eqsecnum]{revtex4}
\usepackage{hyperref}

\usepackage{amsmath}
\usepackage{amsfonts}
\usepackage[dvipdfm]{graphicx}
\usepackage{epstopdf}
\usepackage{subfigure}
\usepackage{epsfig}
\newcommand{\beq}{\begin{equation}}
\newcommand{\eeq}{\end{equation}}

\begin{document}
\setlength{\parindent}{0pt}
\title{Model for Bidirectional Movement of Cytoplasmic Dynein}
\author{S. Sumathy}
\author{S.V.M. Satyanarayana}
\address{
Department of Physics, Pondicherry University \\
R.Venkataraman Nagar, Kalapet,\\
Puducherry 605 014, India. \\
svmsatya@gmail.com}
\date{\today}
\maketitle
\textbf{\large Abstract}\\
\par Cytoplasmic dynein exhibits a directional processive movement on microtubule filaments and is known to move in steps of varying length based on the number of ATP molecules bound to it and the load that it carries. It is experimentally observed that dynein takes occasional backward steps and the frequency of such backward steps increases as the load approaches the stall force. Using a stochastic process model, we investigate the bidirectional movement of single head of a dynein motor. The probability for backward step is implemented based on Crook's fluctuation theorem of non-equilibrium statistical mechanics. We find that the movement of dynein motor is characterized with negative velocity implying backward motion beyond stall force. We observe that the motor moves backward for super stall forces by hydrolyzing the ATP exactly the same way as it does while moving forward for sub stall forces.\\


\section{Introduction}
Molecular motors are nano machines that do work by harnessing chemical energy\cite{dc}. Cytoplasmic dynein is a minus end directed motor protein that moves on microtubule. It is implicated in intracellular transport of vesicles, mRNA, protein complexes etc., from cell cortex to center of the cell ~\cite{aa}. Dynein is a homo-dimer with each head consisting of a ring of six domains\cite{ab}. Four of the six domains of the ring have sites with affinity for ATP binding. Dynein motor exhibits a gear like mechanism in controlling step size in response to the load force\cite{ac}. A model of single head of a dynein motor is simulated by Monte Carlo method\cite{ad}. Singh et. al. computed force velocity relation, step size distribution from the simulation data and studied the ATP dependence of average velocity of the motor.  Further, the load dependence of the step size and ATP concentration dependence of the stall force are obtained by another model with a weak coupling between two reactions coordinates corresponding to chemical reactions and translocation of the motor\cite{ae}. A complete mechanochemical model for hand over hand stepping model of a homodimeric dynein is developed where the ATP hydrolysis cycle is coupled to coarse grained structural model\cite{af,ag}. Dynein motor at different spatio-temporal resolution is studied by multi-scale modeling\cite{ah}. Stochastic process modeling of unidirectional movement of single head of a dynein motor is carried out systematically with one, two and three step process\cite{ai}. While dynein moves processively toward the minus end of the microtubule, it is observed that dynein takes backward steps once in a while and the frequency of such backward steps increases as the load increases\cite{ak}.
\par In the present work, we develop a stochastic process model, along the same lines of the model of Sutapa Mukherji\cite{ai} for bidirectional movement of Cytoplasmic dynein's single head. We use Crook's fluctuation theorem to define the ratio of probability of forward step to that of backward step, as used for kinesin motor\cite{ao,ap}. Dynein can take 8, 16, 24 or 32 nm step sizes depending on the load and ATP concentration. In this study we investigate bidirectional movement of dynein's single head for all the four step sizes.
\par The paper is organized as follows. Description of stochastic process model, typical reaction scheme, different rate constants used in the study, a system of stochastic rate equations and the procedure to compute average velocity of the motor are presented in section 2. Section 3 consists of results and discussion.
\section{Model}

	Microtubule is modeled as a passive one dimensional lattice on which single head of dynein moves. We assume that $dynein^{'}s$ head is always attached to the microtubule site Single head of dynein consists of one primary and three secondary ATP binding sites. It is assumed that out of four ATP binding sites, ATP hydrolysis takes place only at primary site. However, the step length depends on the occupancy of the three secondary sites. If we denote the minimum step size dynein can take as $a$, where $a=8nm$, the step size when n$(0\leq n\leq3)$ secondary sites are occupied is $4a/(n+1)$.
\par If we consider one primary and one secondary sites, the stochastic variable $S_j^{kl}$  denotes the probability that a $dynein^{'}s$ head is on the $j^{th}$ lattice site of microtubule with $k, l = {0, 1}$. Here $k=0(1)$ and $l=0(1)$ signify the corresponding primary or the secondary site being unoccupied (occupied) respectively. In this case, maximum step size of a single head of dynein is 2a and hence we call this a 2a model. If we have two and three secondary ATP binding sites  considered, we refer to them as 3a model and 4a model respectively and denote the corresponding stochastic variable as $S_{j}^{klm}$ and $S_{j}^{klmp}$. Since each index in the superscript of the stochastic variable can be 0 or 1 corresponding to the ATP binding site being unoccupied or occupied respectively, we have number of stochastic variables as 4, 8 and 16 for 2a, 3a and 4a models.
\par The rate of ATP binding on the $i^{th}$ binding site of $dynein^{'}s$ head is denoted as $k_{oni}$ and the rate of ATP unbinding from the same site is denoted as $k_{offi}$. Hydrolysis rate at the primary site depends on the number of ATP occupied in the secondary sites as well as the load that it carries. The load dependence of the hydrolysis rate is given by the following expression
\begin{equation}
k_{cat,i}=A(i)k_{cat,0}\exp[-\alpha Fd(i)/k_{B}T]
\end{equation}
Here, i=1, 2, 3 and 4 and $d(i)=i\times a$, $k_{cat,0}$ is the hydrolysis rate for no load, $\alpha$ is taken as positive since the hydrolysis rate should decrease with the increase in opposing external load and $A(i)=1$ if any of the secondary sites are occupied and is 0.01 if all the secondary sites are unoccupied.
\par Along the lines of previous studies ~\cite{ad,ai}, we assume that the ATP binding rates of secondary sites depend on the load as follows
\begin{equation}
k_{on,2-4}(F)=k_{on,2-4}\exp[Fd_{0}/k_{B}T]
\end{equation}
$d_{0}$ is an adjustable parameter in units of length.
\par Molecular motors are known to take backward steps amidst their processive forward motion. The probability ratio for forward and backward steps for kinesin motor is estimated from experiment ~\cite{am,an}. Subsequently, the expression for the ratio of forward and backward step probabilities was derived from $Crook^{'}s$ fluctuation theorem of non equilibrium thermodynamics for kinesin motor ~\cite{ao,ap}. In this work, we assume that the same expression is valid for dynein motor as well. The ratio of the probability for a forward step to backward step is given as
\begin{equation}
\frac{P_{F}}{P_{B}}=\exp\left[\frac{d_0}{2k_{B}T}[F_{s}-F]\right]
\end{equation}
where $F_S$ is the stall force. Different rate constants and parameters used in the present study have
been taken from previous  studies ~\cite{ad,ai}. They are given in table I.
\begin{table}[ht]
\caption{Rate constants and parameters used in the study.}
  \centering
  \begin{tabular}{|c|c|}
   \hline
    Symbol & Value \\
    \hline
  $K_{B}T$ & 4.1 pN nm \\
  $k_{off1}$ & $10s^{-1}$ \\
  $k_{off2}$& $250s^{-1}$ \\
  $k_{off3}$ & $250s^{-1}$ \\
  $k_{off4}$ & $250s^{-1}$ \\
  $k_{on1}$ & $4\times 10^{5}M^{-1}s^{-1}$ $[ATP]$\\
  $k_{on2}(F=0)$ & $4\times 10^{5}M^{-1}s^{-1}$ $[ATP]$\\
  $k_{on3}(F=0)$ & $k_{on2}(F=0)/4$ \\
  $k_{on4}(F=0)$ & $k_{on2}(F=0)/6$ \\
  $d_0$ & 6nm \\
  $k_{cat,0}$ & $55s^{-1}$ \\
  $\alpha$ & 0.3 \\
  \hline
  \end{tabular}
\end{table}
\par In this study, we have computed the average velocity of single head of dynein molecule with 2a, 3a and 4a models with and without including backward steps. We present the reaction scheme and stochastic rate equation model for bidirectional 3a model as a representative case.
\subsection{The Model}
In a 3a model, we have eight state variables for single head for dynein motor. The state variable vector on the $j^{th}$ lattice site of microtubule is given by
\begin{equation}
\rho_j=[S_j^{000}S_{j}^{001}S_{j}^{010}S_{j}^{100}S_{j}^{011}S_{j}^{101}S_{j}^{110}S_{j}^{111}]^T
\end{equation}
 The reaction scheme for the 3a model is presented in fig 1.
  \begin{figure}
  \includegraphics[width=0.8\textwidth]{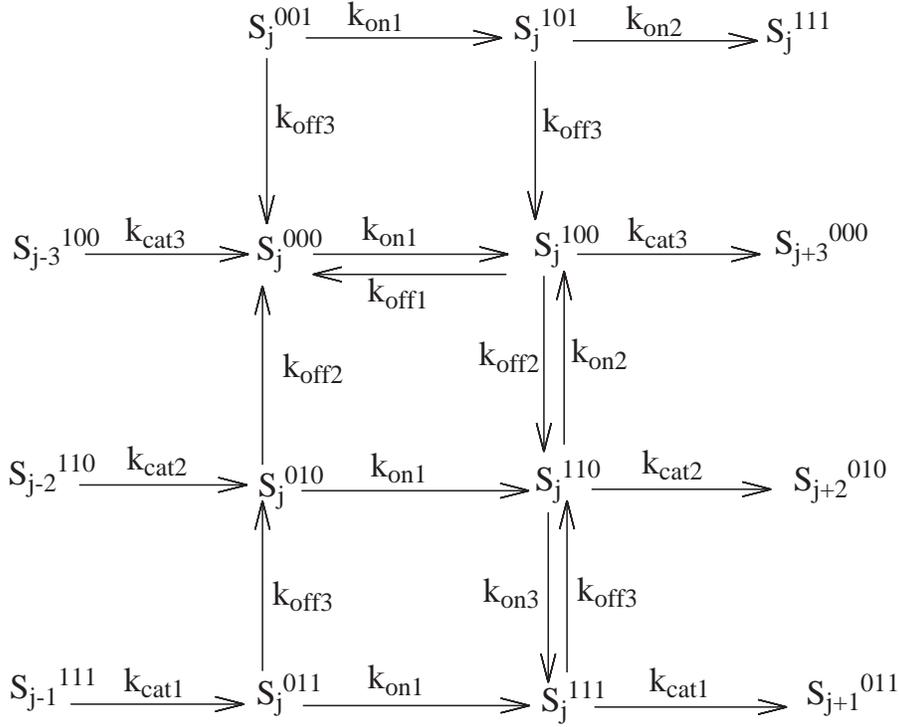}\\
 \caption{Reaction Scheme for $3a$ model of single head of a dynein motor}\label{fig1}
\end{figure}

The stochastic rate equations for the eight state variables of 3a model corresponding to the reaction scheme presented in fig.\ref{fig1} are given below
\begin{eqnarray}
\nonumber
  \frac{dS_{j}^{000}}{dt} &=& -k_{on1}S_{j}^{000}+k_{off3}S_{j}^{001}+k_{off2}S_{j}^{010}+k_{off1}S_{j}^{100}+[P_{F}S_{j-3}^{100}+P_{B}S_{j+3}^{100}]k_{cat3} \\
  \nonumber
  \frac{dS_{j}^{001}}{dt} &=& -(k_{off3}+k_{on1})S_{j}^{001} \\
  \nonumber
  \frac{dS_{j}^{010}}{dt} &=& -(k_{on1}+k_{off2})S_{j}^{010}+k_{off3}S_{j}^{011}+[P_{F}S_{j-2}^{110}+P_{B}S_{j+2}^{110}]k_{cat2} \\
  \nonumber
  \frac{dS_{j}^{100}}{dt} &=& k_{on1}S_{j}^{000}-(k_{cat3}+k_{on2}+k_{off1})S_{j}^{100}+k_{off3}S_{j}^{101}+k_{off2}S_{j}^{110} \\
  \frac{dS_{j}^{011}}{dt} &=& -(k_{on1}+k_{off3})S_{j}^{011}+[P_{F}S_{j-1}^{111}+P_{B}S_{j+1}^{111}]k_{cat1} \\
  \nonumber
  \frac{dS_{j}^{101}}{dt} &=& k_{on1}S_{j}^{001}-(k_{on2}+k_{off3})S_{j}^{101} \\
  \nonumber
  \frac{dS_{j}^{110}}{dt} &=& k_{on1}S_{j}^{010}+k_{on2}S_{j}^{100}-(k_{cat2}+k_{on3}+k_{off2})S_{j}^{110}+k_{off3}S_{j}^{111} \\
  \nonumber
  \frac{dS_{j}^{111}}{dt} &=& k_{on1}S_{j}^{011}+k_{on2}S_{j}^{101}+k_{on3}S_{j}^{110}-(k_{cat1}+k_{off3})S_{j}^{111}
  \nonumber
\end{eqnarray}
In matrix form these equations can be recast as
\begin{equation}
\frac{d \rho}{dt}=[A]\rho_{j}+[B]\rho_{j-1}+[C]\rho_{j+1}+[D]\rho_{j-2}+[E]\rho_{j+2}+[F]\rho_{j-3}+[H]\rho_{j+3}
\end{equation}
Here A, B, C, D, E, F and H are matrices whose elements are various rate constants. Using a generating function, $G(\zeta,t)=\sum_{j=-\infty}^{\infty}\zeta^{j}\rho_{j}$, the above equation can be recast as
\begin{equation}
 \frac{d}{dt}G(\zeta,t)=([A]+\zeta[B]+\frac{1}{\zeta}[c]+\zeta^{2}[D]+\frac{1}{\zeta^{2}}[E]+\zeta^{3}[F]+\frac{1}{\zeta^{3}}[H])G(\zeta,t)=[R(\zeta)]G(\zeta,t)
\end{equation}
where the matrix $R(\zeta)$ is given by
\begin{equation}
{
[R(\zeta)]=\begin{pmatrix} -k_{on1} &k_{off3} & k_{off2} &k_{1} &0 &0 &0 &0\\0 &-(k_{7}) &0 &0 &0 &0 &0 &0\\ 0 &0 &-(k_{6}) &0 &k_{off3} &0 &k_{2} &0\\ k_{on1} &0 &0 &-(k_{4}) &0 &k_{off3} &k_{off2} &0\\ 0 &0 &0 &0 &-(k_{7}) &0 &0 &k_{3}\\ 0 &k_{on1} &0 &0 &0 &-(k_{on2}+k_{off3}) &0 &0\\ 0 &0 &k_{on1} &k_{on2} &0 &0 &-(k_{5}) &k_{off3}\\ 0 &0 &0 &0 &k_{on1} &k_{on2} &k_{on3} &-(k_{cat1}+k_{off3}) \end{pmatrix}
}
\end{equation}
Here $k_{1}=k_{off1}+[\frac{1+\alpha(\zeta^{6}-1)}{\zeta^{3}}]k_{cat3}$, $k_{2}=[\frac{1+\alpha(\zeta^{4}-1)}{\zeta^{2}}]k_{cat2}$, $k_{3}=[\frac{1+\alpha(\zeta^{2}-1)}{\zeta}]k_{cat1}$, $k_{4}=k_{off1}+k_{on2}+k_{cat3}$, $k_{5}=k_{cat2}+k_{on3}+k_{off2}$, $k_{6}=k_{on1}+k_{off2}$ and $k_{7}=k_{on1}+k_{off3}$.
The transition matrix $R[\zeta=1]$ has the property that the sum of all elements in a column is zero. Thus, the largest eigenvalue of the matrix $R[\zeta=1]$ is zero.
\par Our aim is to find the average velocity and it is found from the relation
\begin{equation}
<v>=a\frac{<j>}{t}=a\lambda_{l}^{'}(1)
\end{equation}
Where $\lambda_{l}(\zeta)$ is the largest eigenvalue of $[R(\zeta)]$ and the primes denotes the derivatives of $\lambda$ with respect to $\zeta$. For finding the largest eigenvalue, it is necessary to solve the characteristic equation
\begin{equation}\label{det}
Det[R(\zeta)-\lambda I]=0
\end{equation}
For finding the derivative of the  largest eigenvalue at $\zeta=1$, it is convenient to substitute $\zeta=1+\delta$ and $\lambda=\delta \lambda^{'}(1)+\frac{\delta^{2}}{2}\lambda^{''}(1)$ in Eq.(\ref{det}) and $\lambda^{'}(1)$ is find out by equating the coefficients of $\delta$ to zero\cite{aq,ar}.
\section{Results and Discussion}
Force velocity relation V(F) is one of the important characteristics of the motor. We present in fig. 2, the average velocity of unidirectional movement (corresponding to $P_{F}$ = 1 in the model) of single head of dynein motor in 3a model. The force velocity relation for ATP concentration close to and higher than physiological concentrations (1mM) is found to be
\begin{equation}\label{fv}
V_{U}(F)=V_{U}(0)\exp(-\gamma F)
\end{equation}
\begin{figure}
  \includegraphics[width=0.8\textwidth]{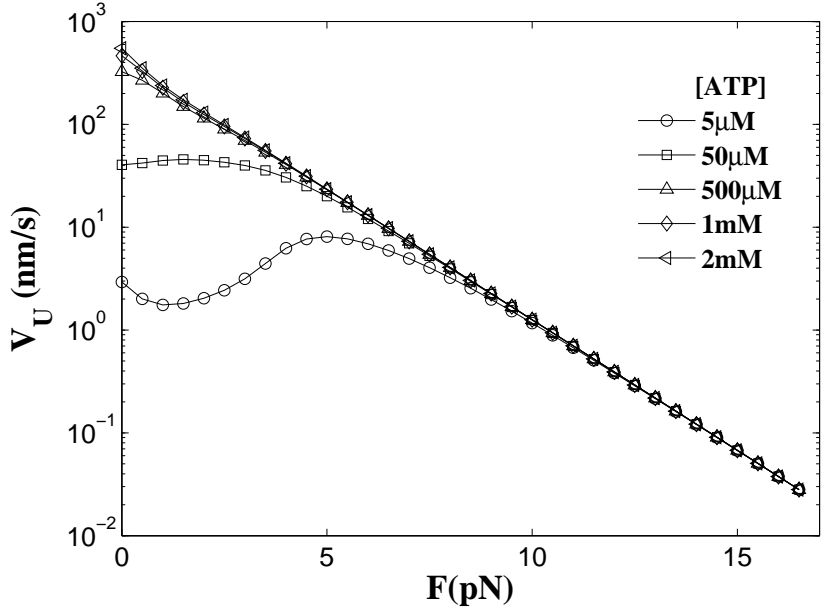}\\
 \caption{Force velocity relation for unidirectional movement of single head of a dynein motor in $3a$ model}\label{fig2}
\end{figure}

From the computed data of force velocity curves, we estimated the parameters $V_{U} (0) = 425.2 nm/s$ and $\gamma = 0.583pN^{-1}$. The observed force velocity relation in Eq.(\ref{fv})  is found to differ from the general form of force velocity relation for motors\cite{as}.\\
\begin{figure}
  \includegraphics[width=0.8\textwidth]{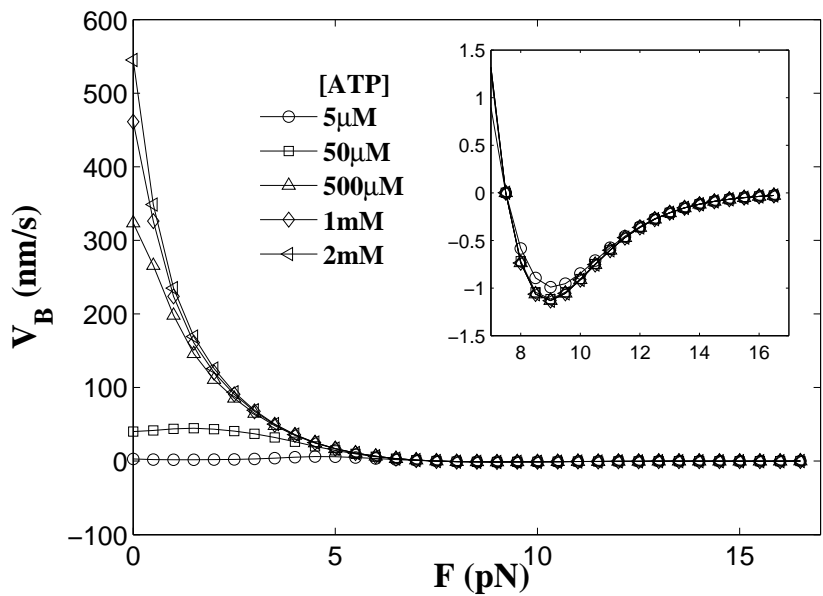}\\
 \caption{Average velocity versus force for bidirectional movement of single head of a
dynein motor for different ATP concentrations. Inset shows the average velocity beyond stall
force (7.5 pN).}\label{fig3}
\end{figure}

Average velocity for a bidirectional motion of single head of dynein in 3a model as a function of force for different ATP concentrations is shown in fig.\ref{fig3}. It can be seen from the fig. 3 that at the stall force, the average velocity of the motor is zero. This is because the probability of forward step is equal to the probability of backward step, that is, on an average number of forward steps is equal to number of backward steps. This is in good agreement with observations made in single molecule experiments with force feedback optical tweezers at an ATP concentration of 1mM\cite{ak}. It can be seen from the inset of fig.\ref{fig3} that beyond stall force (7.5 pN) the velocity is negative, that is the motor walks backwards (in a direction opposite to its natural direction of motion in the absence of any load force). We observe from fig.3 and the inset of fig.\ref{fig3} that the average negative velocity as a function of force for super stall forces is independent of ATP concentration. This result is also in a good agreement with single molecule experiments\cite{ak}. Further, as F increases beyond stall force, the negative average velocity decreases and approaches zero for high super stall force.\\
\begin{figure}
  \includegraphics[width=0.8\textwidth]{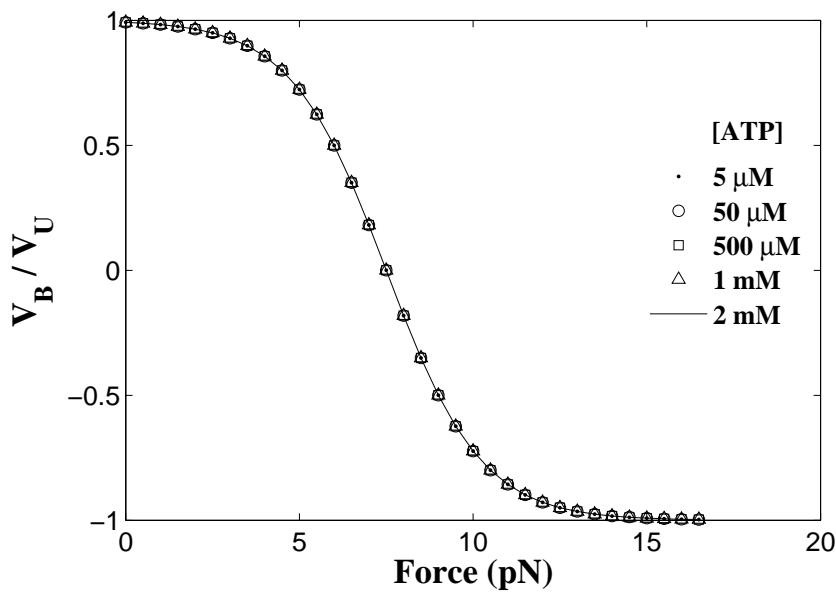}\\
 \caption{Ratio of velocities corresponding to bidirectional and unidirectional movement
of dynein motor as a function of force for different ATP concentrations.}\label{fig4}
\end{figure}

\par Figure \ref{fig4} represents the ratio of velocities of bidirectional and unidirectional movement. It can be observed that this ratio is independent of ATP concentration in a range of ATP concentrations from $5\mu M$ to 10 mM studied in this work for all forces. As $F\rightarrow 0$, $V_{B} = V_{U}$, as can be expected. For $F >> F_S$,it is observed that $V_{B} = -V_{U}$. This implies for $F >> F_S$, the cargo acts like a rigid wall and the motor gets reflected with the same velocity in the opposite direction, in the average sense.The ratio given in fig.\ref{fig4} has the following functional form
\begin{equation}
\frac{V_{B}}{V_{U}}=\tanh \left(\frac{(F_{s}-F)\delta}{2k_{B}T}\right)
\end{equation}
\par The value of $\delta$ is found to be 3 nm. For sub stall forces, bidirectional velocity is a fraction of unidirectional velocity. This reduction in velocity is due to number of backward steps that the motor head takes in a bidirectional motion. For super stall forces, the number of backward steps is more than the number of forward steps and as a result the average velocity is negative, with a magnitude equal to a  fraction of the unidirectional velocity. For $F >> F_S$, almost all steps are backward steps. However, we observe that the average negative velocity of the bidirectional motor decreases in much the same way as the average positive velocity of the unidirectional motor approaching zero from negative and positive directions respectively. This can be understood as follows. Dynein motor hydrolyses ATP and the rates of ATP binding to different binding sites and ATP hydrolysis depend on the force. For every ATP hydrolysis event, a step forward or backward is taken based on probabilities obtained from fluctuation theorem. For super stall forces, the rates corresponding to ATP binding and hydrolysis are low such that number of ATP hydrolysis events become very less and tend to zero. The average bidirectional velocity is small negative for $F >> F_S$ implies that ATP hydrolysis events, however small they are, lead to backward steps of the motor.  During the backward movement the motor hydrolyzes ATP in exactly the same way as it does while moving forward for force $F < F_S$.
\par We observe that the 4a model where we have four ATP binding sites for single head of the dynein gives similar results. However, the magnitude of the average velocity is higher for 4a model compared to 3a model in both unidirectional as well as bidirectional cases. For example, the velocity of bidirectional movement of single head of dynein motor under no load is 461 nm/s in 3a model and 711 nm/s in 4a model. The velocity of dynein motor for 4a model is in reasonable agreement with experimental observations ~\cite{at}.
\par In conclusion, we have modeled the unidirectional and bidirectional movement of single head of dynein motor using stochastic rate equations. Backward steps are implemented using $Crook^{'}s$ fluctuation theorem of nonequilibrium statistical mechanics. We model single head of dynein with one primary and one, two and three secondary ATP binding sites and presented the results for one primary and two secondary ATP binding sites, the 3a model. We find the magnitude of unidirectional or bidirectional velocity is larger when more number of secondary ATP binding sites are considered for low load forces. The computed velocity from 4a model of dynein motor is found to be in good agreement with measured velocities. We find that the motor moves backwards for super stall forces. The ratio of velocities corresponding to bidirectional movement and unidirectional movement exhibits an ATP concentration independent universal behavior. We find that for super stall forces, the motor moves backwards by hydrolyzing ATP exactly the same way as it does while moving forward for sub stall forces.

\
\end{document}